\documentclass[preprint,12pt,authoryear]{elsarticle}

\usepackage{amssymb}
\usepackage[version=4]{mhchem}

\usepackage{multirow}
\usepackage{color, soul}

\journal{Chemical Physics Letters}

\DeclareUnicodeCharacter{2212}{-}

\begin{document}

\begin{frontmatter}



\title{Electron scattering from aminoacetonitrile: effects of polarisation-correlation and basis-set on cross section}


\author[inst1]{Irabati Chakraborty}

\affiliation[inst1]{organization={Department of Physics, Indian Institute of Technology (Indian School of Mines), Atomic and Molecular Physics Laboratory},
            city={Dhanbad},
            postcode={826004}, 
            state={Jharkhand},
            country={India}}

\author[inst1]{Bobby Antony\corref{cor1}}
\ead{bobby@iitism.ac.in}

\begin{abstract}
Aminoacetonitrile occupies a prime importance in the interface between astrochemistry and prebiotic chemistry. Its detection in the ISM establishes it as part of the organic inventory of star-forming regions, while its role as a glycine precursor highlights its significance for origins-of-life scenarios. In this work, electron scattering from aminoacetonitrile has been studied using the $R$-matrix method in the low-energy range from $\sim$0 to 10 eV. The calculations were carried out within the $C_{s}$ point group using static-exchange (SE), static-exchange plus polarisation (SEP), and configuration interaction (CI) models, with two basis sets (6-311G* and cc-pVTZ) to understand their dependence on cross section. Various scattering observables, such as differential elastic cross section, integral elastic, excitation, and momentum transfer cross sections, were examined. Since aminoacetonitrile is a prebiotically relevant molecule, these findings provide valuable insight into electron-driven processes in complex organic systems and form a theoretical foundation for future work on electron-induced reactivity in prebiotic and astrophysical environments.
\end{abstract}
 
\begin{graphicalabstract}
\includegraphics[scale=0.55]{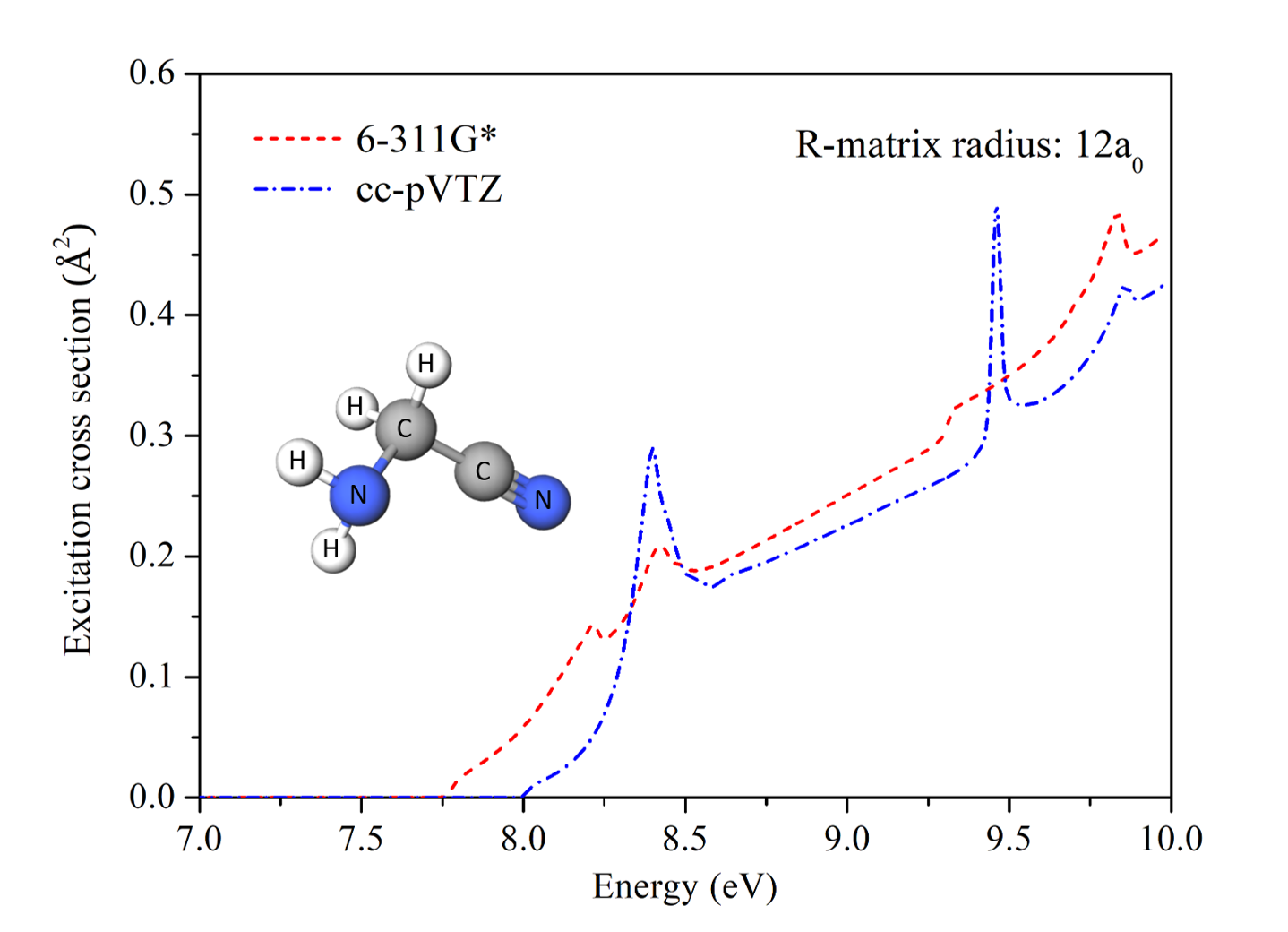}
\end{graphicalabstract}

\begin{highlights}
\item Low-energy (below 10 eV) electron scattering from the aminoacetonitrile molecule is investigated here.
\item Sensitivity studies were conducted employing three models (SE, SEP, and CI) and two different basis sets (6-311G* and cc-pVTZ) under the R-matrix method.
\item Differential elastic and integral elastic, excitation, and momentum transfer cross sections are computed on a fine energy grid.
\item Resonance features are observed, implying temporary negative ion formation during collision.
\item Results provide valuable insight into electron-driven processes in prebiotic and astrophysical environments.
\end{highlights}


\end{frontmatter}




\section{Introduction}

Aminoacetonitrile, \ce{NH2CH2CN}, is a key molecule in astrochemistry since it is closely related to glycine, the simplest amino acid. Its first confirmed detection in space was reported toward the hot molecular core Sagittarius B2(N) using the IRAM 30\,m telescope, where Belloche and colleagues identified many rotational transitions and determined its abundance in this chemically rich region \citep{belloche2008detection}. More recently, aminoacetonitrile has also been observed in the hot molecular core G10.47+0.03 with ALMA, where it was found at temperatures of about 120 K and with abundances consistent with chemical models of hot cores \citep{manna2022identification}. These results indicate that aminoacetonitrile is not unique to the Galactic Center but can be produced in different high-mass star-forming regions. Laboratory spectroscopy has advanced in parallel. Detailed far-infrared and rotational studies have improved line catalogues and made it possible to detect vibrationally excited states of aminoacetonitrile in space \citep{melosso2020far, jiang2023insights}. This progress ensures reliable identification of the molecule in crowded astronomical spectra.
\begin{figure}[h]
\centering
  \includegraphics[scale=1]{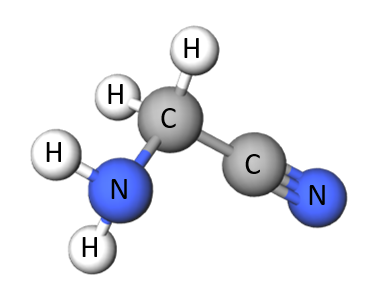}
  \caption{Structure of aminoacetonitrile}
  \label{fig1}
\end{figure}

The importance of aminoacetonitrile lies in its prebiotic chemistry. In the classical Strecker synthesis, glycine is formed from ammonia, formaldehyde, and hydrogen cyanide via aminoacetonitrile as an intermediate \citep{koch2008theoretical}. Theoretical studies show that this chemistry can occur on icy dust grains in cold interstellar environments, producing aminoacetonitrile under astrophysical conditions \citep{koch2008theoretical}. Laboratory experiments confirm that both ultraviolet irradiation and simple thermal processing of ices containing ammonia and nitriles produce aminoacetonitrile \citep{danger2011experimental, danger2011experimentalb}. Additional work has shown that aminoacetonitrile can form in water-rich ices and remains stable under moderate heating \citep{borget2012aminoacetonitrile, noble2013thermal}. As glycine is difficult to detect directly in the interstellar medium \citep{snyder2005rigorous, lattelais2011detectability}, aminoacetonitrile serves as a valuable alternative for understanding how far interstellar chemistry proceeds toward amino acid formation.

Recent studies have highlighted the role of energetic processing in this chemistry. Low-energy electrons (LEEs, $\leq 20$ eV), created when cosmic rays ionise interstellar gas and ices, are especially important. They interact strongly with organic molecules, often through resonant processes that can lead either to fragmentation or to the production of reactive radicals \citep{boyer2016role, wu2023role, boudaiffa2000resonant, mason2014electron}. In aminoacetonitrile, the presence of both an amino and a nitrile group introduces multiple reactive centres where LEEs may attach, potentially leading to selective bond cleavage, radical formation, or intramolecular rearrangements. Such processes could contribute both to the destruction of aminoacetonitrile in the ISM and to the generation of reactive intermediates that feed into further prebiotic chemistry. From a theoretical and experimental standpoint, studying electron scattering from aminoacetonitrile provides an opportunity to investigate how multifunctional prebiotic molecules respond to electron-driven processes. Unlike smaller nitriles (such as acetonitrile) or simple amines, aminoacetonitrile combines two chemically distinct functional groups, whose interplay may give rise to complex resonance structures and scattering behaviour. Understanding these processes is critical for assessing the molecule’s stability under astrophysical conditions and for evaluating its prebiotic potential.

For molecules similar to aminoacetonitrile, such as acetonitrile, low-energy electron experiments have revealed strong resonances and efficient dissociative electron attachment channels \citep{sailer2003low, maioli2017low, zawadzki2018low}. Existing work on aminoacetonitrile focuses on dissociative electron attachment (DEA). Pelc \emph{et al.} measured anion yields following low-energy electron collisions with aminoacetonitrile \citep{pelc2016formation}. These results show that LEEs can both destroy aminoacetonitrile and generate reactive fragments that may feed into further chemistry. Despite the progress, data on electron scattering and electron-induced processes from aminoacetonitrile are still limited. To date, no elastic or electronic excitation cross sections in the low energy range, i.e., below the ionisation threshold, have been reported. Measurements of elastic and inelastic cross sections, resonance lifetimes, and fragmentation branching ratios are essential for refining astrochemical models. Without them, predictions of aminoacetonitrile abundances and lifetimes remain uncertain. Because the molecule contains both amino and nitrile groups, it is expected to support resonance channels that control its chemistry under electron impact. Thus, studying low-energy electron interactions with aminoacetonitrile is important not only for understanding its observed abundances but also for clarifying its role as a prebiotic precursor in interstellar, cometary, and planetary environments.

\section{Theory}

The UK R-matrix method \citep{tennyson2010electron}, implemented through the Quantemol-N software \citep{tennyson2007quantemol}, was used in this work to calculate low-energy electron scattering cross sections. The R-matrix approach is a well-established framework for studying electron–molecule collisions. It works by dividing configuration space into two regions separated by a spherical boundary of radius \(a\), placed at the center of mass of the target molecule.

Inside the sphere, known as the inner region, the molecule and the incoming electron are treated together as an \((N+1)\) electron system. In this region, short-range effects such as electron exchange and correlation dominate, so the system behaves like a bound state. Outside the sphere (outer region), the scattered electron no longer feels these short-range interactions. Instead, it interacts only with the long-range multipole potential of the target molecule. Here, the wavefunction of the scattering electron is expressed in a form that reduces the problem to a system of coupled differential equations, which can then be solved with standard numerical techniques.

The radius of the R-matrix sphere must be chosen carefully so that the short-range forces vanish at the boundary. In this study, a radius of \(12 a_0\) was used. The Hamiltonian of the \((N+1)\) electron system is built separately for the inner and outer regions, and the solutions in these two regions are matched at the boundary of the sphere to ensure continuity of the wavefunctions.

The wavefunction in the inner region is represented using the close-coupling approximation:
\begin{equation}
\psi^{N+1}_k
= A \sum_{i j} a_{i j k}\, \Phi_i(x_1, x_2, \ldots, x_N)\, u_{i j}(x_{N+1})
\;+\; \sum_i b_{i k}\, \chi_i(x_1, x_2, \ldots, x_{N+1}),
\end{equation}
where \(A\) is the antisymmetrization operator. In this expression, \(\Phi_i\) are the wavefunctions of the target molecule, while \(u_{i j}\) are continuum orbitals representing the scattering electron. The coefficients \(a_{i j k}\) give the weight of each target–continuum combination in forming the inner-region solution. The terms \(\chi_i\) are square-integrable (\(L^2\)) functions confined within a finite region, and the coefficients \(b_{i k}\) describe their contribution to the solution.

From the inner-region eigenfunctions, the R-matrix is constructed at the boundary. This R-matrix is then propagated outward and matched with asymptotic solutions obtained using the Gailitis expansion \citep{gailitis1976new}. Through this procedure, the \(K\)-matrix is determined. Using the POLYDCS code \citep{sanna1998differential}, the \(K\)-matrix is converted into the \(S\)- and \(T\)-matrices, which are directly related to the scattering observables. The scattering cross sections are finally calculated from the elements of the \(T\)-matrix.

To examine the scattering under different levels of approximation, three models were used: the static exchange (SE), static exchange plus polarisation (SEP), and configuration interaction (CI) models. The SE model provides a simple baseline, where the target molecule is represented by a frozen Hartree–Fock potential. The SEP model improves on this by including the distortion of the target’s electron cloud due to the incident electron, thus accounting for polarisation effects. The CI model gives the most accurate description by also including excitations of the target, which are essential for capturing resonance effects and providing reliable excitation dynamics.

\subsection{Present Target State}

The point group for aminoacetonitrile is $C_{s}$. For the present calculation, the geometrical parameters for ground state aminoacetonitrile have been taken from the CCCBDB database \citep{CCCBDB}. The basis sets used to construct the target wavefunctions are 6-311G* and cc-pVTZ. The scattering calculations are done using static exchange (SE), static exchange plus polarisation (SEP), and configuration interaction (CI) models. The ground state electronic configuration is given as:  
\[
1a'^{2}, \; 2a'^{2}, \; 3a'^{2}, \; 4a'^{2}, \; 5a'^{2}, \; 6a'^{2}, \; 7a'^{2}, \; 8a'^{2}, \; 1a''^{2}, \; 9a'^{2}, \; 2a''^{2}, \; 10a'^{2}, \; 11a'^{2}, \; 3a''^{2}, \; 12a'^{2}.
\]

Additionally, 5 virtual orbitals are included in the CI calculation to represent the electronic states of the molecule beyond the ground state. Out of these 30 electrons, 20 core electrons are frozen and do not take part in the excitation. The remaining 10 electrons in the outermost orbitals take part in the excitation. Hence, the active space is composed of the 5 ground-state outermost orbitals and 5 virtual orbitals, containing 10 electrons, given as  
\(
(10a', \; 11a', \; 12a', \; 13a', \; 14a', \; 15a', \; 2a'', \; 3a'', \; 4a'', \; 5a'')^{10}.
\)
For the CI calculation, a total of 9772 configuration state functions (CSFs) are generated for the 25 target states.

\section{Results and Discussions}

This section depicts the results obtained for electron-aminoacetonitrile scattering in the energy range $\sim$0 to 10 eV. The results are presented in graphical formats. 

\subsection{Eigenphase sum}

\begin{figure}[h]
\centering
  \includegraphics[scale=0.15]{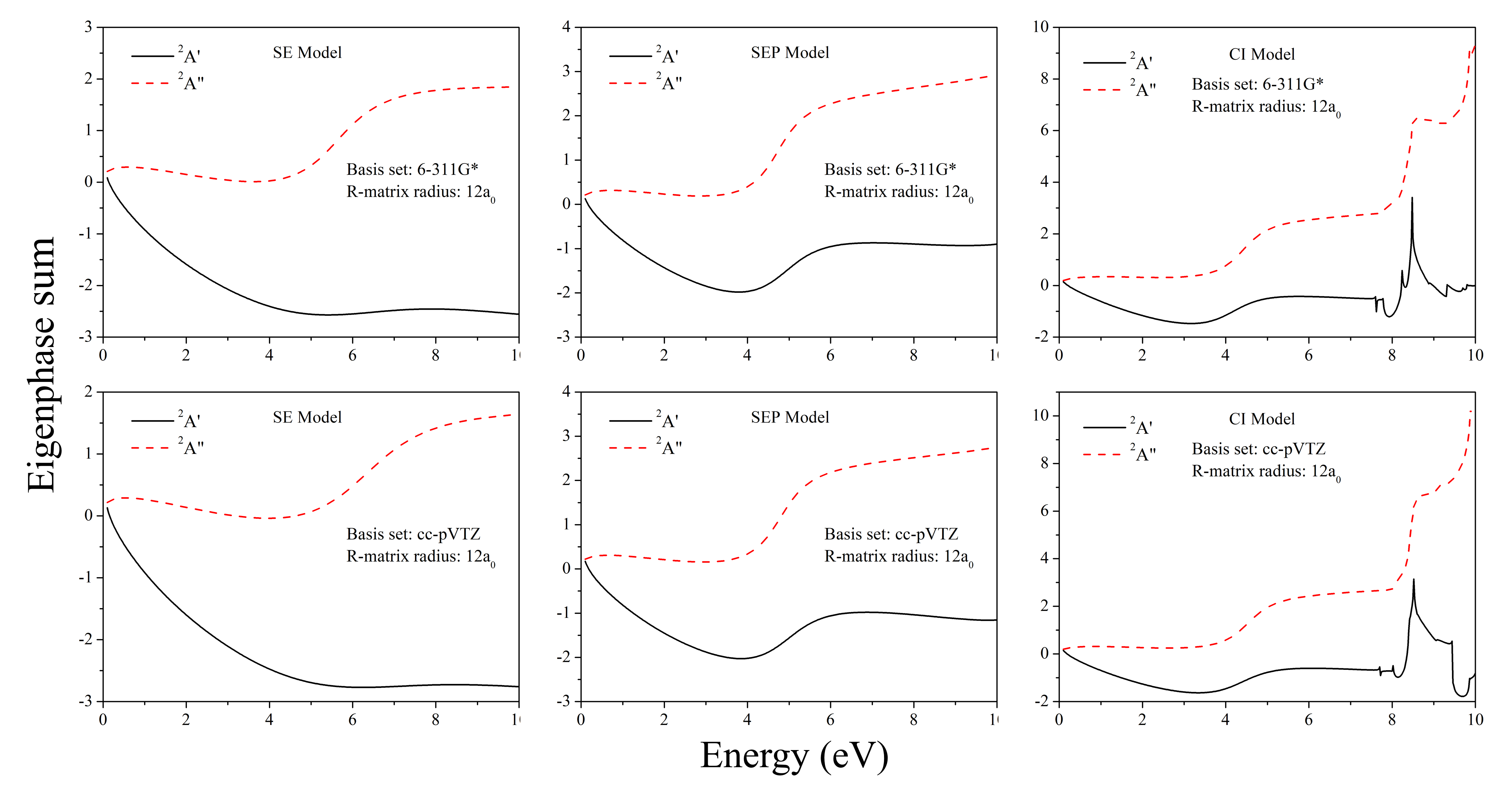}
  \caption{Eigenphase sum of aminoacetonitrile for different symmetries calculated using SE, SEP, and CI models}
  \label{fig2}
\end{figure}

The eigenphase sums for electron scattering from aminoacetonitrile in the $C_{s}$ point group were obtained within the $R$-matrix framework for two irreducible representations, $^{2}A'$ and $^{2}A''$. Calculations were performed with three target models: static-exchange (SE), static-exchange plus polarisation (SEP), and configuration interaction (CI) using two different basis sets (6-311G* and cc-pVTZ), with an $R$-matrix radius of $12 \, a_{0}$. The results are presented in figure \ref{fig2}. 

In the static-exchange (SE) approximation, the scattering eigenphases are relatively smooth, with only weak evidence for sharp resonance behaviour. For the 6-311G* basis, a single broad structure was identified in the $^{2}A''$ channel at 5.70 eV with a width of 1.98 eV. The corresponding eigenphase sum shows only a gradual rise, lacking the steep $\pi$-like jump characteristic of well-defined resonances. With the cc-pVTZ basis, a similar feature is found slightly higher in energy at 6.45 eV with a broader width of 2.71 eV. No clear resonant structure was detected in the $^{2}A'$ symmetry for either basis set at the SE level. Overall, the SE model yields only diffuse, basis-dependent structures without strong eigenphase support for genuine resonances. 

When target polarisation is included in the static-exchange-plus-polarisation (SEP) model, the situation changes noticeably. For 6-311G*, two resonances emerge: a $^{2}A'$ state at 5.09 eV with a width of 1.80 eV and a $^{2}A''$ state at 4.75 eV with a narrower width of 1.20 eV. The eigenphase sums in this case show sharper, arctangent-like rises around these energies, confirming their resonant nature. Using the cc-pVTZ basis yields nearly identical results: the $^{2}A'$ resonance appears at 5.15 eV with a width of 1.84 eV, while the $^{2}A''$ resonance is located at 4.81 eV with a width of 1.25 eV. Thus, polarisation stabilises and lowers the resonance energies compared to SE and produces consistent results across basis sets. 

The configuration-interaction (CI) model introduces explicit correlation and provides the clearest and most reliable resonance picture. With 6-311G*, two low-lying resonances are found: a $^{2}A'$ state at 4.29 eV with a width of 
1.59 eV and a $^{2}A''$ state at 4.43 eV with a width of 1.09 eV. Both are supported by steep eigenphase jumps, consistent with true temporary anion formation. In addition, a very narrow higher-lying $^{2}A'$ resonance is fitted at 7.62 eV with a width of $3.25 \times 10^{-3}$ eV. For the cc-pVTZ basis, the same low-energy features are reproduced with remarkable consistency: the $^{2}A'$ and $^{2}A''$ resonances appear at 4.53 eV and 4.55 eV, with widths of 1.69 eV and 1.12 eV, respectively. At higher energies, two additional very narrow resonances are observed in $^{2}A'$: one at 7.71 eV with a width of $3.50 \times 10^{-3}$ eV and another at 9.94 eV with an extremely small width of $1.02 \times 10^{-4}$ eV. These sharp eigenphase features could represent weakly coupled Feshbach-type states, but their small residues and high sensitivity to model details warrant caution in interpretation. 

Overall, the progression across models and basis sets is consistent. SE provides only broad, poorly defined features; SEP lowers and sharpens the first resonances into the 4–5 eV region with improved eigenphase support; and CI yields robust low-energy resonances at $\sim$4.3–4.6 eV in both symmetries, with widths of $\sim$1–1.7 eV. The larger cc-pVTZ basis generally shifts SE resonances upward and produces marginally narrower SEP and CI widths, but the low-energy CI resonances converge very well between the two basis sets. The additional high-energy, long-lived CI resonances at 7–10 eV appear in both basis sets, though with slight variations in energy and width. These features are highly sensitive and may correspond either to real narrow states or to artifacts of the fitting, requiring further validation through direct inspection of eigenphase sums and cross sections.

\subsection{Elastic cross section}

\begin{figure}[h]
\centering
  \includegraphics[scale=0.2]{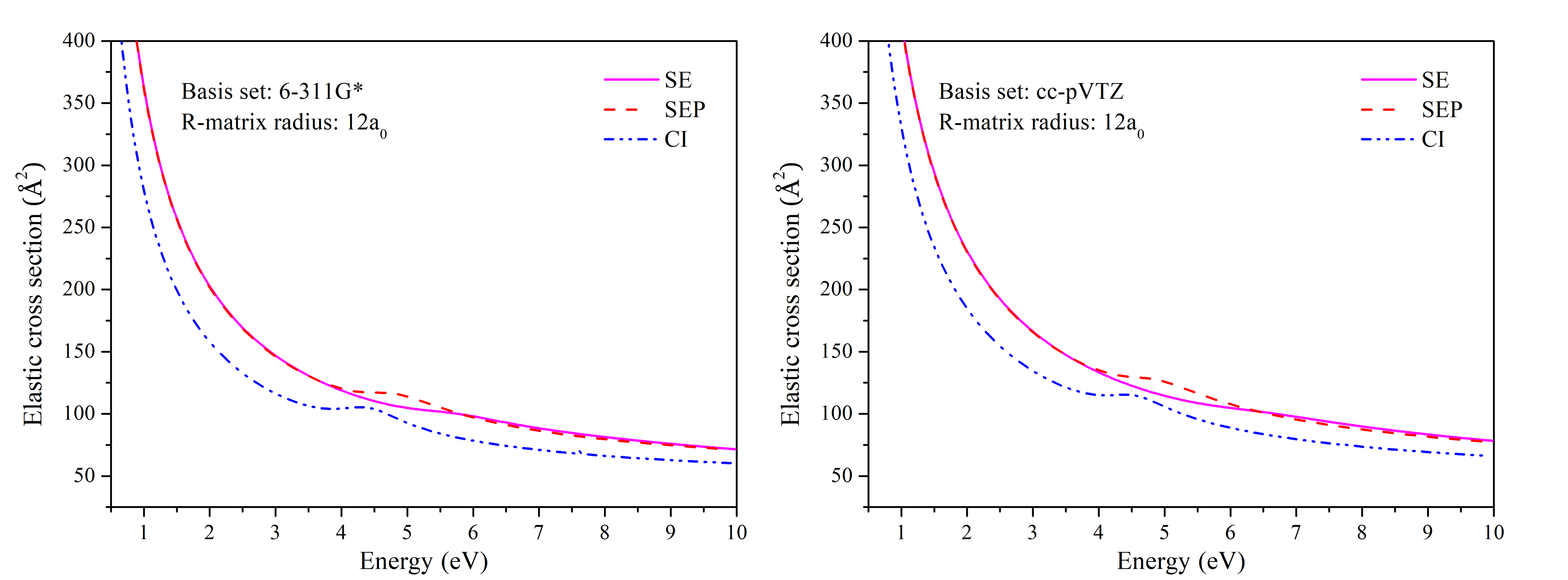}
  \caption{Elastic cross section of aminoacetonitrile}
  \label{fig3}
\end{figure}

Figure \ref{fig3} represents the integral elastic electron scattering cross sections of aminoacetonitrile for various models and basis sets. The results show systematic dependence on both the chosen basis set and the treatment of electronic correlation. For both the 6-311G* and cc-pVTZ basis sets, the cross sections display the characteristic steep rise at low energies, consistent with the long-range polarisation interaction, followed by a gradual decrease with increasing electron energy. This overall trend reflects the dominance of $s$-wave scattering near the threshold. 

The most pronounced model dependence occurs in the 4–5 eV region, where the inclusion of polarisation and correlation modifies both the shape and magnitude of the cross section. At the static-exchange (SE) level the elastic cross section is smooth. This behaviour is consistent with the eigenphase sums, which showed only weak, broad features in the SE model without clear resonant jumps. In contrast, when target polarisation is included in the SEP model, the cross sections exhibit a distinct resonant enhancement near 4.8–5.1 eV. This structure is observed for both basis 
sets, although it is slightly more pronounced in the cc-pVTZ calculation and corresponds directly to the low-lying resonances identified in the eigenphase analysis. 

At the configuration-interaction (CI) level, the eigenphase analysis revealed robust resonances in both symmetries at 4.3–4.6 eV with widths of about 1–1.7 eV. In the elastic cross sections these resonances manifest as modest shoulders or plateaux, rather than sharp peaks. The overall magnitude of the CI cross sections in this energy region is reduced compared to SE and SEP, which suggests that correlation effects redistribute scattering flux in a way that sharpens the eigenphase behaviour but does not strongly increase the total elastic signal. This is a common effect in polyatomic systems, where correlation improves the resonance description but also couples elastic and inelastic channels, thereby lowering the integrated elastic yield. 

The comparison between basis sets shows that 6-311G* and cc-pVTZ give very similar elastic cross sections once correlation is included. At the SEP level, cc-pVTZ yields a slightly more pronounced resonant shoulder near 5 eV, while at the CI level, both basis sets produce nearly identical curves across the full energy range. This agreement indicates that the CI/cc-pVTZ model provides the most reliable description of the low-energy resonances and the corresponding elastic scattering behaviour. The narrow high-energy resonances seen in the CI eigenphases between 7 and 10 eV do not contribute prominently to the total elastic cross section, consistent with their very small fitted widths and likely weak coupling. The convergence of the two basis sets at the CI level reinforces the robustness of these low-energy resonances, which are therefore assigned as genuine scattering states of aminoacetonitrile.

\subsection{Excitation cross section}

\begin{figure}[h]
\centering
  \includegraphics[scale=0.2]{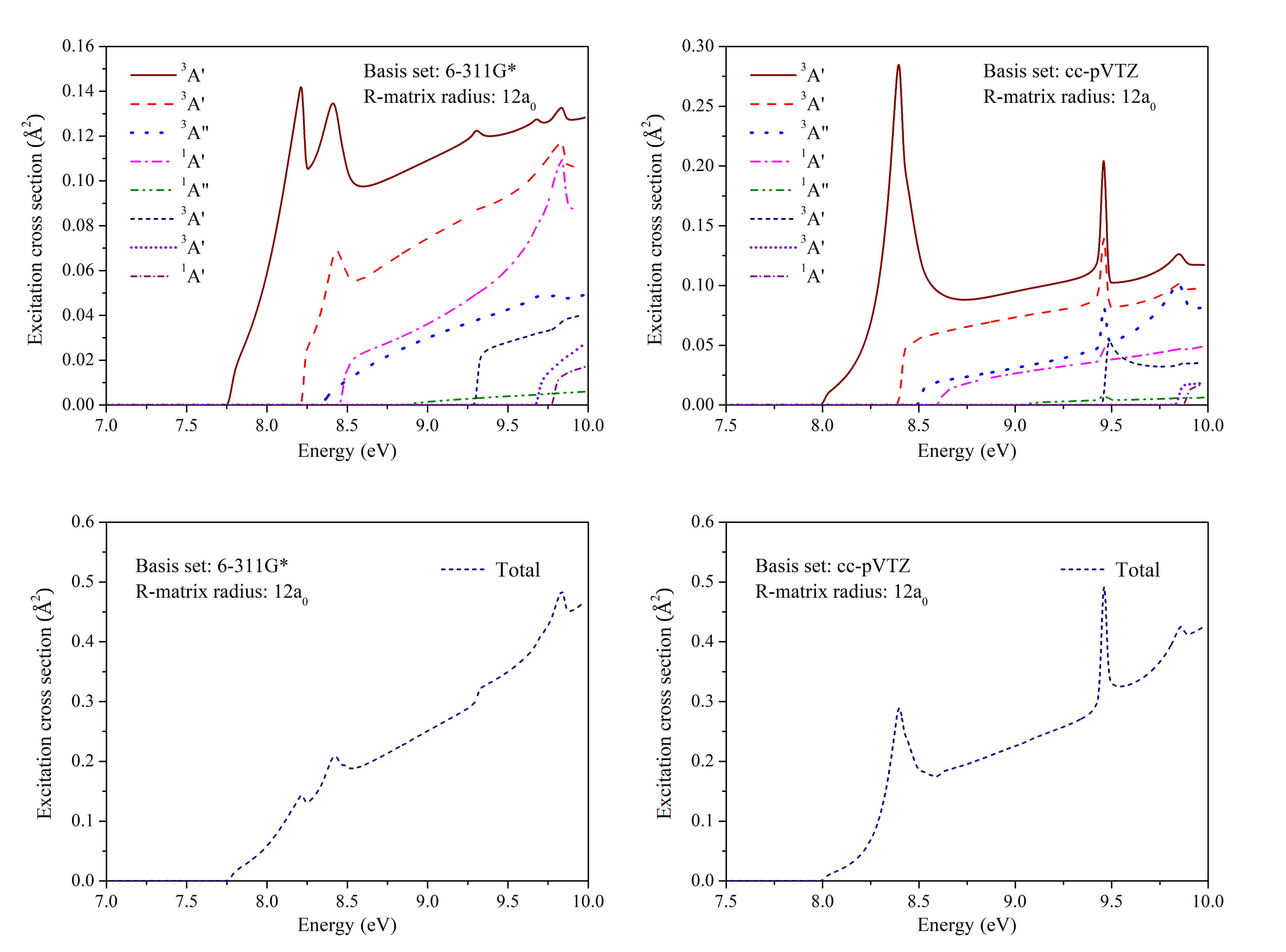}
  \caption{Excitation cross section of aminoacetonitrile}
  \label{fig4}
\end{figure}

The electronic excitation cross sections of aminoacetonitrile into the lowest singlet and triplet electronic states were calculated for both the 6-311G* and cc-pVTZ basis sets, and the results are presented in figure \ref{fig4}. The graphs exhibit several distinct threshold behaviours and resonance-enhanced features between 7 and 10 eV. The excitation processes are symmetry-resolved in the upper panels and summed in the lower panels for the two basis sets. 

For the 6-311G* basis, the excitation cross sections are relatively small in magnitude, with most channels contributing less than $0.15 \, \text{\AA}^{2}$. The dominant contribution arises from the triplet $^{3}A'$ state, which shows a broad increase above threshold and weak resonant modulations around 8.5–9.5 eV. Other channels, such as $^{3}A''$ and $^{1}A'$, contribute smaller stepwise increases at higher energies. The total excitation cross section for 6-311G* rises gradually with energy, reaching about $0.45 \, \text{\AA}^{2}$ at 10 eV, and shows several modest shoulders that reflect the threshold of individual target states. 

The cc-pVTZ basis yields significantly larger excitation cross sections and sharper resonance structures. The most prominent feature is a strong resonance in the $^{3}A'$ channel, peaking near 8.5 eV with a maximum of 
$\sim 0.28 \, \text{\AA}^{2}$. A second enhancement occurs at $\sim 9.5$ eV in both $^{3}A'$ and $^{3}A''$ channels, producing a sharp peak in the total excitation cross section that exceeds $0.45 \, \text{\AA}^{2}$. Compared with 6-311G*, the cc-pVTZ results clearly demonstrate that a more extensive basis set stabilises excited states, lowers thresholds, and enhances coupling to the scattering electron, thereby producing stronger and more structured excitation signals. 

The total excitation cross sections further emphasise the difference between the two basis sets. With 6-311G*, the overall magnitude is lower, and the resonance features appear as small bumps. In contrast, with cc-pVTZ, the total cross section rises more steeply above threshold, exhibits a pronounced peak at 8.5 eV, and shows an additional sharp maximum at 9.5 eV. These peaks align with the narrow high-energy resonances identified in the eigenphase analysis at 7–10 eV, supporting their interpretation as genuine, long-lived scattering states. This indicates that the more diffuse and flexible cc-pVTZ functions capture the coupling to excited states more accurately, leading to a clearer resonance signature. The presence of such strong features in the cc-pVTZ results also highlights the sensitivity of excitation channels to basis set quality, with 6-311G* underestimating resonance strengths and over-smoothing the excitation profile.

\subsection{Momentum transfer cross section}

\begin{figure}[h]
\centering
  \includegraphics[scale=0.2]{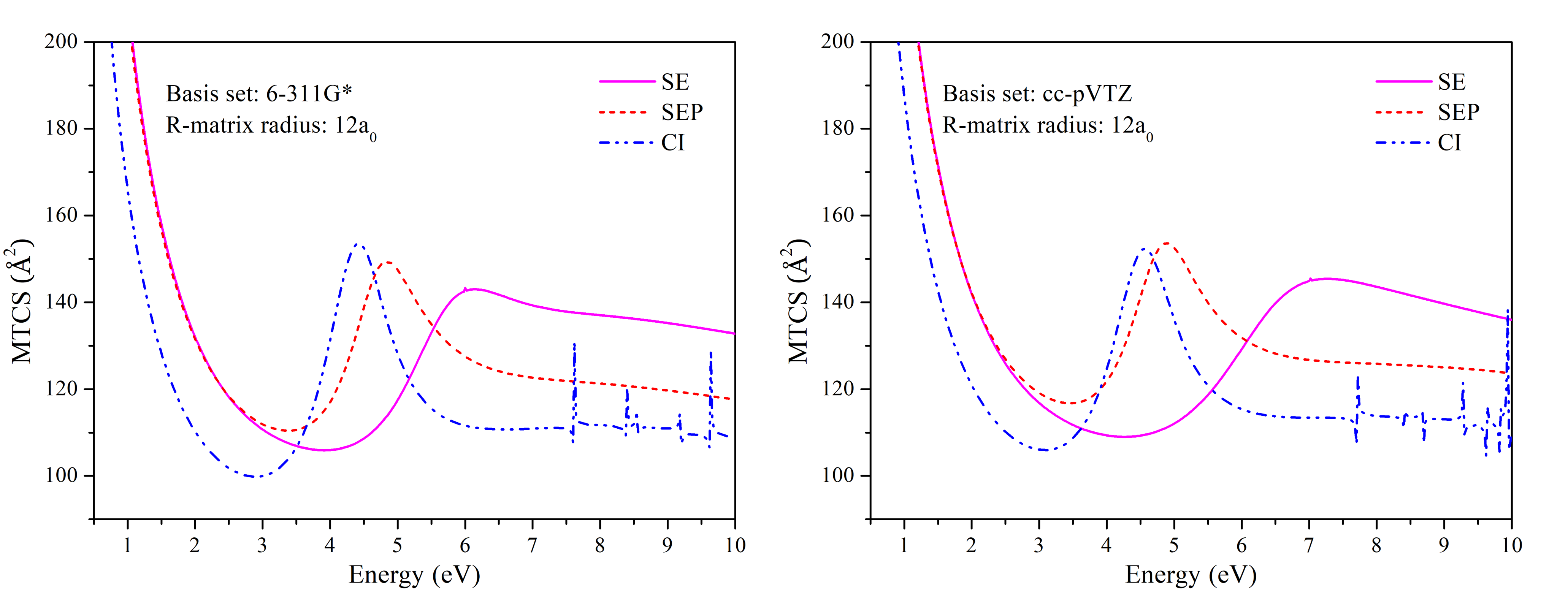}
  \caption{Momentum transfer cross section of aminoacetonitrile}
  \label{fig5}
\end{figure}

The calculated momentum transfer cross sections (MTCS) for aminoacetonitrile are presented in figure \ref{fig5} for the two basis sets, comparing the static-exchange (SE), static-exchange plus polarisation (SEP), and configuration interaction (CI) models. At low energies, below 2 eV, all models yield very large MTCS values, consistent with strong forward scattering driven by the long-range dipole and polarisation potential. As the energy increases, the MTCS values decrease sharply, reaching a pronounced minimum near 3 eV. Beyond this minimum, the MTCS exhibits resonance-driven enhancements in the 4–6 eV region. For the 6-311G* basis set, the SEP and CI models show clear peaks around 4.5–5 eV, while the SE model produces a broader rise centred near 6 eV. The cc-pVTZ results show a similar trend, with the SEP and CI models 
exhibiting peaks near 4.5–5 eV, reflecting the stabilisation of the shape resonance identified in the eigenphase sum analysis. The SE model again produces a broader and less well-structured enhancement, consistent with its neglect of polarisation effects. 

At higher energies, from 7 to 10 eV, the MTCS values flatten to 110 - 130~$\text{\AA}^{2}$, with the CI models consistently predicting smaller cross sections than SE and SEP. The reduction in MTCS in CI models reflects more accurate cancellation between resonance contributions and background scattering, leading to weaker net momentum transfer. Additionally, the oscillatory behaviour visible in the CI curves at higher energies corresponds to Feshbach resonances, reflecting coupling to inelastic channels that open in this energy region, in agreement with the excitation cross section results. 

Overall, the MTCS results demonstrate the strong influence of target polarisation and electron correlation effects. The SEP and CI models consistently reduce the magnitude relative to SE while introducing structure associated with low-lying resonances. These results, taken together with eigenphase, elastic, and excitation data, demonstrate that a careful treatment of correlation and basis-set convergence is essential for predicting momentum transfer processes, which are particularly relevant for modelling electron-driven transport and energy deposition in molecular environments such as interstellar ices or prebiotic gas mixtures.

\subsection{Differential cross section}

\begin{figure}[h]
\centering
  \includegraphics[scale=0.195]{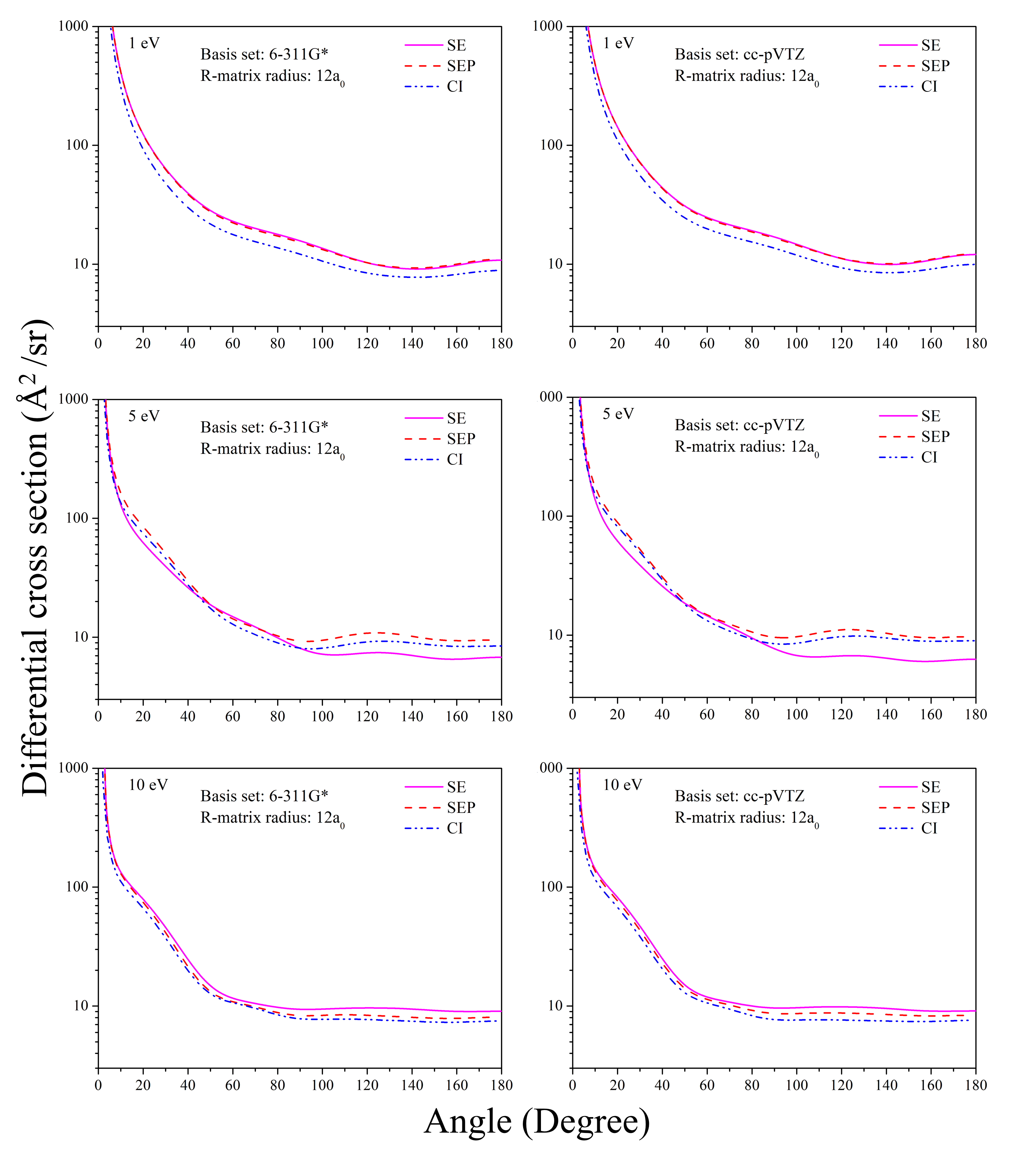}
  \caption{Differential cross section of aminoacetonitrile}
  \label{fig6}
\end{figure}

Figure \ref{fig6} presents the differential cross sections (DCS) for elastic electron scattering from aminoacetonitrile, calculated at 1, 5, and 10 eV using the SE, SEP, and CI models with both basis sets (6-311G* and cc-pVTZ). At all energies the angular distributions are strongly forward-peaked, reflecting the dominance of long-range interactions at small scattering angles. The inclusion of polarisation (SEP) and correlation (CI) generally lowers the DCS compared with SE, especially at intermediate and large scattering angles, consistent with enhanced absorption of flux into polarisation and correlation channels.

At 1 and 10 eV the three models show a very similar angular dependence, with only modest quantitative differences between SE, SEP, and CI. However, at 5 eV a distinct behaviour emerges: while SEP and CI reduce the forward scattering relative to SE, at scattering angles beyond $90^{\circ}$ both SEP and CI exceed the SE result. This backward enhancement is not present at 1 or 10 eV, indicating it is tied to structure in the 4–6 eV energy region. The most plausible explanation is resonance-related: eigenphase analysis places low-lying resonances in this interval (between 4.5 and 5 eV), and inclusion of polarisation and correlation modifies partial-wave phase shifts so that higher-order contributions interfere constructively at large angles. This behaviour demonstrates the sensitivity of the angular distributions to resonance dynamics and the importance of polarisation effects in describing backscattering. 

Overall, the DCS results confirm the predominance of forward scattering in aminoacetonitrile, while also highlighting the role of resonance–polarisation coupling in shaping the angular distributions at intermediate energies.

\section{Conclusion}

This work presents a detailed $R$-matrix study of electron scattering from aminoacetonitrile in the $\sim$0–10 eV range, reporting integral elastic, excitation, momentum transfer, and differential elastic cross sections. By systematically comparing static-exchange (SE), static-exchange plus polarisation (SEP), and configuration interaction (CI) models with two basis sets (6-311G* and cc-pVTZ), the role of polarisation, correlation, and basis-set choice in shaping scattering dynamics has been established.

The eigenphase analysis revealed a clear progression across models. SE calculations produced only diffuse, basis-dependent features without strong evidence of resonances. The inclusion of polarisation in the SEP model stabilised and sharpened resonances, consistently locating them in the 4–5 eV region. The CI model provided the most reliable description, yielding robust $^{2}A'$ and $^{2}A''$ resonances at $\sim$4.3–4.6 eV with widths of 1–1.7 eV, well reproduced across both basis sets. At higher energies (7–10 eV), extremely narrow resonances were detected in CI, suggesting possible long-lived Feshbach-type states.

Elastic and momentum-transfer cross sections supported these resonance assignments. While SE produced smooth, featureless behaviour, SEP and CI introduced structures near 4–5 eV that aligned with the low-energy eigenphase resonances. The momentum transfer cross sections further emphasised resonance-driven enhancements in the 4–6 eV region and showed overall reductions in magnitude when correlation was included. Excitation cross sections highlighted the importance of basis-set quality. With 6-311G*, excitation probabilities were underestimated, and resonance structures appeared smoothed out. In contrast, cc-pVTZ yielded significantly larger cross sections with pronounced peaks at 8.5 and 9.5 eV, consistent with the narrow high-energy resonances seen in eigenphase analysis. This underscores that diffuse basis functions are essential for accurately describing coupling to excited states. Differential cross sections confirmed forward-peaked scattering dominated by long-range interactions but also revealed resonance-driven modifications at 5 eV, particularly enhanced backscattering when polarisation and correlation were included.

In summary, this study demonstrates that reliable predictions of electron–aminoacetonitrile scattering require the explicit inclusion of polarisation and correlation effects, supported by a sufficiently flexible basis set. The robust low-energy resonances at $\sim$4.5 eV are assigned as genuine scattering states and may facilitate temporary anion formation in environments where aminoacetonitrile is present, such as interstellar ices or prebiotic atmospheres. The additional high-energy resonances at 7–10 eV suggest further resonance-driven pathways for electronic excitation. The CI model with the cc-pVTZ basis set emerges as the most consistent and accurate framework, providing a solid foundation for future investigations into the role of electron attachment and excitation processes in the chemical evolution of prebiotic molecules.

\appendix


\bibliographystyle{elsarticle-harv} 
\bibliography{cas-refs}





\end{document}